\documentclass{ws-ijmpd}

\begin{document}

\markboth{Michael Stamatikos}
{GRB Astrophysics in the Swift Era and Beyond}

%
\catchline{}{}{}{}{}
%

\title{GRB ASTROPHYSICS IN THE SWIFT ERA AND BEYOND\footnote{Contributed to the Proceedings of the 2nd Heidelberg Workshop: High-Energy Gamma-rays and Neutrinos from Extra-Galactic Sources (Max Planck Institute for Nuclear Physics).}}

\author{MICHAEL STAMATIKOS}
\address{Center for Cosmology and Astro-Particle Physics (CCAPP) Fellow/Department of Physics, \\The Ohio State University, 191 West Woodruff Avenue, Columbus, OH 43210, USA
\\Michael.Stamatikos-1@nasa.gov}

\maketitle


\begin{abstract}
Gamma-ray Bursts (GRBs) are relativistic cosmological beacons of transient high energy radiation whose afterglows span the electromagnetic spectrum. Theoretical expectations of correlated neutrino emission position GRBs at an astrophysical nexus for a metamorphosis in our understanding of the Cosmos. This new dawn in the era of experimental (particle) astrophysics and cosmology is afforded by current facilities enabling the novel astronomy of high energy neutrinos, in concert with unprecedented electromagnetic coverage. In that regard, GRBs represent a compelling scientific theme that may facilitate fundamental breakthroughs in the context of Swift, Fermi and IceCube. Scientific synergy will be achieved by leveraging the combined sensitivity of contemporaneous ground-based and satellite observatories, thus optimizing their collective discovery potential. Hence, the advent of GRB multi-messenger astronomy may cement an explicit connection to fundamental physics, via nascent cosmic windows, throughout the next decade.
\end{abstract}


\section{Electromagnetic Emission: The Synergy of Swift and Fermi}

The Swift MIDEX explorer mission\cite{Gehrels:2007}, comprised of the wide-field ($\sim1.4$ sr, half-coded) hard X-ray (15-150 keV) Burst Alert Telescope (BAT), and the narrow-field (0.2-10 keV) X-Ray (XRT) and (170-600 nm) Ultraviolet-Optical (UVOT) Telescopes, has revolutionized our understanding of GRBs. The intrinsic multi-wavelength instrumentation, coupled with a rapid ($\lesssim$ 100 seconds) autonomous slew capability, has ushered in an unprecedented era of source localization precision $\left(\lesssim1^{\prime}-4^{\prime}\right)$ that is disseminated in real-time ($\lesssim10$ seconds) via the GRB Coordinate Network (GCN)\cite{Barthelmy:2008}, thus spear-heading international ground-based and satellite multi-wavelength follow-up campaigns. Swift's unique dynamic response and spatial localization precision, in conjunction with the aforementioned correlative ground-based follow-up efforts, have resulted in redshift determinations for $\gtrsim133$ GRBs, including the most distant cosmological explosion, GRB 080913 at $z=6.695\pm0.025$\cite{Greiner}, which has begun to constrain progenitor models\cite{Belczynski}. Selection effects, such as detector composition and long accumulation timescales, bias BAT towards long, soft GRBs with lower characteristic photon energy $\left(\text{E}_{\text{peak}}\right)$. Consequently, BAT GRBs comprise a separate statistical class, as is demonstrated by their fluence and redshift distributions $\left(\bar{z}\approx2.3\right)$, from classical Burst and Transient Source Experiment (BATSE) GRBs. High-z GRBs afford an unprecedented opportunity to probe the earliest epoch of stellar formation via the detection of Population I, II (normal) and III (massive) stars. Meanwhile, the association of GRB 050509B with an elliptical host galaxy\cite{Gehrels:2005} supports the idea that short GRBs may arise from compact binary mergers such as the coalescence of neutron stars ($\sim1.4$ M$_{\odot}$) with other neutron stars (DNS) or black holes (BH-NS/BH-BH)\cite{Bloom:2006}. With an estimated orbital lifetime of $\sim15$ years, it is anticipated that Swift will continue to operate until $\sim2017$.

\begin{figure}[pb]
\centerline{\psfig{file=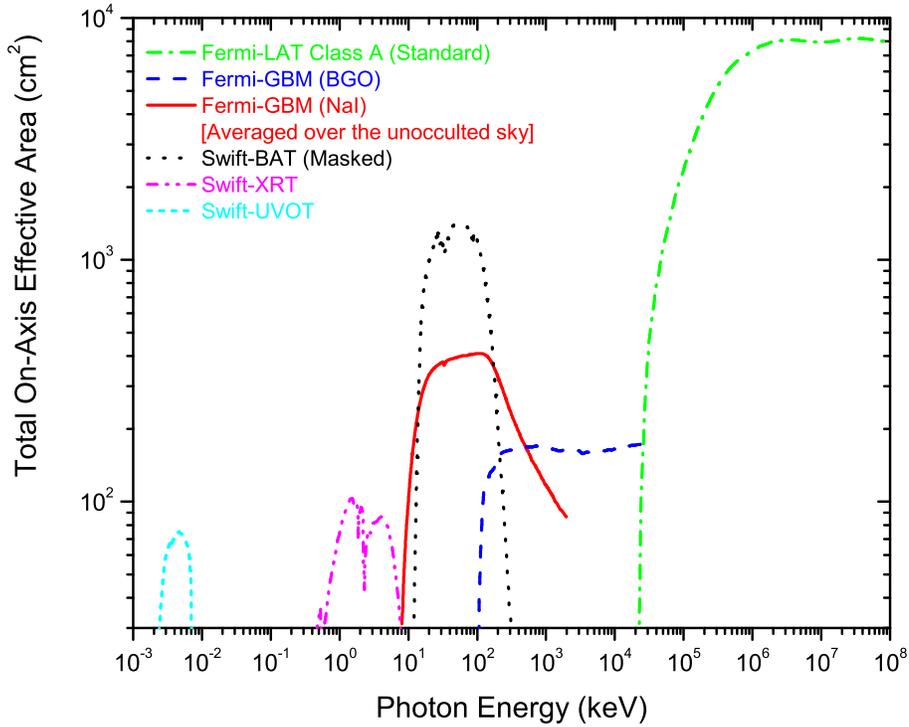,width=12cm}}
\vspace*{8pt}
\caption{Effective areas for Swift-BAT/XRT/UVOT and Fermi-GBM/LAT. Exceptional bursts may be detected over 11 energy decades facilitating temporal and spectral evolution studies.\label{Overlapped_Effective_Areas}}
\end{figure}

Fermi, which is comprised of the ($<$20 MeV to $>$300 GeV) Large Area Telescope (LAT) and the (10 keV - 30 MeV) Gamma-ray Burst Monitor (GBM), launched on June 11, 2008 and has an anticipated mission lifetime of $\sim10$ years, taking it into $\sim2018$. Fermi has already made some remarkable detections\cite{Longo:2009} including the first GeV emission from a short burst (GRB 081024B) as well as the burst with the most isotropic energy release (GRB 080916C). The GBM, consisting of 12 NaI (10-1000 keV) and 2 BGO (0.15-25 MeV) detectors, monitors $\sim8$ steradians of the sky, and, in concert with LAT, enables Fermi to continuously span 7 energy decades, but not at the same sensitivity. As illustrated in Figure~\ref{Overlapped_Effective_Areas}, Fermi's effective area drops by over $\sim1.5$ orders of magnitude from LAT (GeV) to GBM-NaI (keV) energies, while the (masked) BAT ($\sim$20-100 keV) low energy effective area surpasses GBM-NaI's by over a factor of $\sim3$. Furthermore, although Swift has detected $\sim400$ GRBs, the majority of $\overline{\text{E}}_{\text{peak}}\sim250$ keV values lie beyond BAT's canonical energy range. Thus, correlated 
observations would simultaneously augment Fermi-GBM's low energy response while increasing the number of $\text{E}_{\text{peaks}}$ for Swift-BAT GRBs. 

Additionally, since Swift's high fidelity localization precision surpasses GBM's by over $\sim2-3$ orders of magnitude, BAT will facilitate GRB follow-up campaigns for Fermi bursts. In this manner, we expect that $\sim35\%$ of bursts within joint BAT-GBM analyses will be accompanied by panchromatic ground-based follow-up observations resulting in spectroscopic redshift determinations and host galaxy identifications. If a LAT GRB triggers Swift, chances favor subsequent detections by it's NFI's, since $\sim95\%$ of BAT GRBs are observed with XRT ($T\lesssim200$ ksec) and $\sim60\%$ have accompanied optical measurements from a combination of UVOT and ground-based observations. Hence, Fermi and Swift are poised to simultaneously span eleven decades in spectral energy (see Figure~\ref{Overlapped_Effective_Areas}), and have already commenced correlated GRB observations for $\sim25$ GRBs. It is anticipated that such joint studies will be possible a few times per month for an annual rate of $\sim32\pm17$ GRBs\cite{Stamatikos:2008f}.

\section{Non-Electromagnetic Emission: A New Window with IceCube}

Thus far, a canonical phenomenological description, known as the fireball model\cite{Meszaros:1993b}, has successfully described most prompt and afterglow GRB electromagnetic observations. However, despite great strides made by satellite-and ground-based electromagnetic observations, the details of GRB progenitor(s) remain concealed. This physical limitation is due to the initial optical thickness of the fireball, prior to adiabatic relativistic expansion, which precludes us from witnessing the genesis of a GRB. The relativistic\footnote{Apparent super-luminal motion has been observed in the radio afterglow of GRB 030329\cite{Taylor:2004}.} nature of GRBs, inferred by the necessity for the expanding fireball shell to be optically thin, requires bulk Lorentz boost factors in the range $100\lesssim\Gamma\lesssim1000$, with typical values of $\sim300$ for isotropic emission geometry. \emph{Since we are fundamentally limited by the optical opacity of the source at the early stages of its inception, perhaps the next evolution in our understanding of the microphysics of the central engine will be realized via the detection of non-electromagnetic emission.}

The observed isotropic and cosmological spatial distribution of GRBs, coupled with the fact that the energy injection rates of ultra high-energy cosmic rays (UHECRs) and GRBs are similar, resulted in the phenomenological suggestion that GRBs may be the sources of UHECRs\cite{Waxman:1995}. Canonical fireball phenomenology, in the context of hadronic (Fermi) acceleration within the astrophysical jet, predicts a taxonomy of correlated MeV to EeV neutrinos from GRBs of varying flavor and arrival times. Ideal for detection are $\sim$ TeV-PeV muon neutrinos\cite{WaxmanBahcall:1997}, which arise as the leptonic decay products of photomeson interactions ($p+\gamma \rightarrow \Delta ^{+} \rightarrow \pi ^{+} + [n] \rightarrow \mu^{+}+\nu_{\mu} \rightarrow e^{+} + \nu_e +$ $\bar{\nu}_{\mu}+\nu_{\mu}$), within the internal shocks of the relativistic fireball. Since the prompt $\gamma$-rays act as the ambient photon target field, these neutrinos are expected to be in spatial and temporal coincidence, which imposes a constraint that is tantamount to nearly background-free searches\cite{Stamatikos:2005b,Stamatikos:2006} in telescopes such as IceCube\cite{Kappes:2009}.

\section{Multi-Messenger GRB Astronomy: Synthesis \& Science Impact}

GRBs are beacons for multi-messenger astronomy\cite{Stamatikos:2009b} that serve as astrophysical laboratories for Special/General Relativity, particle astrophysics and cosmology. In the electromagnetic regime, Swift's dynamic response and localization precision will complement prompt emission from Fermi, while facilitating ground-based follow-up via the GCN. Such unprecedented broad-band electromagnetic capability will enable joint GRB data sets that will enhance our understanding of burst parameter classifications, enable routine determinations of E$_{\text{peak}}$ values, explore GRB emission geometry, and help test the viability of various redshift estimation methods\cite{Stamatikos:2008b}. In addition, a more accurate normalization between prompt and afterglow emissions will facilitate the determination of GRB energy budgets, while enabling investigations of spectral and temporal evolution\cite{Stamatikos:2009} over unprecedented decades of energy.

The detection of non-electromagnetic emission from GRBs, via high-energy neutrinos with IceCube, has the potential for broad scientific breakthroughs. A positive detection of high-energy neutrinos would confirm hadronic acceleration in the relativistic GRB-wind providing critical insight to the associated micro-physics of the fireball while revealing an astrophysical acceleration mechanism for UHECRs, thus resolving a century old enigma. Given such tremendous discovery potential for science synergy and impact, our view of the Cosmos is bound to change forever.

\section*{Acknowledgments}

M. Stamatikos is supported by a CCAPP Fellowship at the Ohio State University.


\end{document}